# Implications of ageing for the design of cognitive interaction systems

*Lucas Morillo-Mendez*

We are living longer in times of the biggest technological revolution humanity had ever seen before. Trying to understand how these two facts interact with each other, or more specifically, trying to maximise the benefits that new developments could potentially offer for the enhancement of the quality of life of older adults, is a task on which we have already begun to work. In particular, the rapid growth of cognitive interaction systems (CISs) – technologies that learn and interact with humans to extend what human and machine could do on their own – offers a promising landscape of possibilities.

The development of assistive technologies has benefited from interdisciplinary collaboration. For example, technologies may be developed for medical reasons between engineers and researchers in medicine. These could include devices that alert family members of dangerous situations in case blood pressure is too high or if the older adult falls or needs any kind of urgent assistance. Other reasons for creating new technologies for senior citizens are more related to psychological wellbeing: social robots for addressing problems such as loneliness and depression are a reality right now (Figure 1). However, there is another possibility that does not imply creating new technologies for assisting older adults, but creating new technologies that everybody could easily interact with, despite their final purpose and independent of the age of the user. Interdisciplinarity still plays a role for this to happen. Designers of new technologies must cooperate with those who provide the guidelines for correctly adapting them to the needs of different kinds of people. Even though we can create technologies specifically designed for helping older adults, they should also be able to enjoy every kind of new technology available for everybody else, such as the Internet, a kitchen robot or an autonomous car. One current example of an existing technology not designed with older adults in mind is videogames. Older adults playing videogames may sound like an alien idea to the reader, but as videogames age, the same thing will happen with their users. In a recent interview, Hamako Mori, an 89-year-old YouTube celebrity and videogame streamer, mentions that she is not playing multiplayer games anymore because she thinks that she slows down other players, but she believes that dedicated servers for older players will be created for everyone to compete in equal conditions (Coverly, 2019). Hamako is a good example of an older adult who wants to enjoy new technologies already available for everybody else. Purely assistive technologies do not have to be the only means of increasing their quality of life.

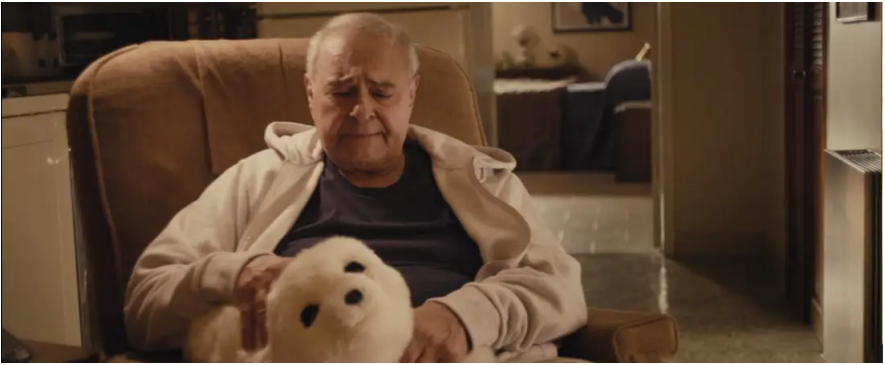

*Figure 1. PARO the seal is a real example of technology for providing companionship to older adults. Source: Master of None series, Netflix (2015).*

## Getting older in the fourth industrial revolution

Despite the number of possibilities ahead of us, the co-occurrence of this elongation in the lifespan and the fourth industrial revolution has a concrete negative aspect. I will briefly try to illustrate it with a personal anecdote related to one of the older adults I personally appreciate most, which happens to be my father. I remember that one time while in elementary school during the 90s, when I was around nine years old, I asked him:

> '-Dad, why don't we have Internet? Some friends have it at home and they can use it for the writings we have to do for school.'

> His answer might sound shocking today:

> '-Why would you want that? We have an encyclopedia if you want to consult any information.'

I would not say that my father was being ignorant, a judgement that can be easily made from our modern perspective. From my point of view, I find it easier to see that the world around himself was already moving too fast, and catching up is not always easy. In that time, my father and many other people could not see how the Internet had already started to make a physical encyclopedia or a newspaper feel analogue or outdated. Luckily, after some time he changed his mind and we got an Internet connection that he is using on a daily basis. This anecdote serves to illustrate something: nowadays, the feeling of the world being constantly ahead of us is permanent and more relevant than ever before, especially from a technological perspective. We are living longer than before while we are subject to a fast-paced rhythm in a world where the feeling of being left behind starts as soon as we blink for a second. If this is hard for all of us, close your eyes and imagine for a moment how complicated it can get for older adults to keep track.

But there are reasons to be optimistic, at least from the technological perspective. We are currently immersed in what has been called the fourth industrial revolution, a period when the appearance of ground-breaking technologies is constantly challenging and moulding our way of living. If we stop and think about how this period could improve the lives of an ageing population and we design new technologies taking into consideration their needs, our future needs, it will be possible that nobody feels left behind. We will try to turn the double-edged sword of a continuously developing strange technology into an inclusion tool. For example, right now the topic of autonomous driving cars is being sold in terms of the 'comfort' of being able to do something other than driving, and in terms of ecology, safety and in terms of power and status (as every expensive car is in reality). Apparently, someone has forgotten that autonomous driving has the potential and of enhancing the independence of people with reduced mobility or cognitive decline. With knowledge and good will, innovative technology adapted for the needs of older people could be designed. In the beginning of the 21$^{st}$ century, we have witnessed the development of some interdisciplinary areas of knowledge concerned with empowering the end user, such as informatics, ergonomics or human-machine interaction (HMI), so there are reasons to believe that technology development companies will consider this knowledge. At the end of the day, extending the range of potential users would also have an impact on their benefits.

**Understanding the cognitive needs of the user**
How could we know what makes older adults benefit the most from different technologies? There is not a simple answer to this question. The first thing that might come to the reader's mind is to ask them directly, and this partially correct.

Asking people their opinion about a technology is the most straightforward way of gathering valuable information, and in fact it is traditionally associated with the world of marketing. Focus groups or interviews are familiar concepts that we associate with testing a product, but on some occasions, for opinions to be valid and solid, the product should be in a post-development phase that allows for it to be tested (or at least it should be somehow familiar to the user). When this is not the case, when opinions are based on predefined ideas, results should be taken with caution: my opinion about autonomous driving might be biased towards the idea that the car will take care of everything and I will be able to take splendid naps, while another person might think that autonomous driving is the synonym of a four-wheeled coffin. While testing it on a safe ground and controlled environment (simulator, virtual reality, closed circuit, etc.) and basing our opinions on that experience is a possibility, this approach implies designing a prototype that might be far from the final product. Even though it is not the ideal, it is hard to deny that this kind of tests may be helpful for the immediate steps of the design.

However, opinions are subjective and there is more valuable information that can be gathered that the person testing the system is not aware of. In a risk assessment experiment, independent of the opinion of how dangerous the participant thinks a certain situation could be, we can gather other external information such as whether the participant was watching the information that announced danger by means of tracking their gaze. Imagine an autonomous car that suddenly disconnects. It is possible that some opinions about this event do not reflect the danger of the situation, and at the same time, there is a possible scenario in which the situation is dangerous indeed. This paradoxical situation may occur if the alert of disconnection is so subtle that the person does not see it, making it a dangerous situation despite subjective opinions. This objective perspective is more related to cognitive psychology and how academics work, but the situations and scenarios used in this type of research might be even further away from real technologies than the prototypes the industry works with. Cognitive psychology, the branch of psychology that studies how humans perceive, interpret and interact with the world, has traditionally used rather artificial scenarios and stimuli. Findings that have been found in a laboratory, using basic shapes and sounds and all kinds of controlled stimuli, are not easily generalisable to specific real tasks because the world is more complex than these settings.

These two ways of gathering information are not mutually exclusive and they can complement each other. Nevertheless, given the advances in new tools of research in cognition (e.g. physiological measures for heart rate or skin temperature, eye-tracking or neuroimage techniques that allow us to correlate internal states with brain activity) and the rapid growth of CISs, the disciplines of cognitive psychology and computer science are right now focused on each other. Because of the novelty and the youth of the HMI field, it is complicated to determine the extent to which previous findings in cognitive psychology are applicable to these new scenarios. The main focus of this research is to apply new paradigms of cognitive psychology in order to design better CISs for older adults. The next section enumerates some characteristics of human cognition and perception with the aim of clarifying the potential impact of cognitive psychology on the design of these new technologies. However, even if it won't be its main focus, this research can also be complemented with qualitative questionnaires to gather opinions. After all, what is the point of creating something we are sure is effective if people just don't like it?

## Some characteristics of cognition

It was previously mentioned that experiments in cognitive psychology have traditionally used rather artificial stimuli and settings. This is not wrong in itself, but generalising results that come from an extremely controlled environment to our everyday life and the future interactions to come is risky. If we are thinking of cognitive psychology as a science applied to real human interactions, we should be able to study cognition in real contexts. Humans assign meaning to their surroundings and integrate information from different sensory modalities; for that reason we say that our perception is **multimodal**. Studying different features from different modalities (i.e. pitch or energy for sound, shape and colour for vision) in isolation is useful for understanding them alone, but humans do not perceive the world as a simple sum of features. Our perception of the world is holistic, it is more than the sum of its parts. For this reason, psychology experiments are starting to use more naturalistic tasks and stimuli that are more relevant to everyday and real interactions (Risko & Kingstone, 2017). For example, eye tracking, a technique that allows us to see the exact points a person is looking at in real time, can be used for exploring what people at looking at in a driving context, as well as many other contexts (Figure 2).

Another reason for insisting on naturalistic research as a need for the design of CISs is related to a theory of cognition called **embodied cognition**. In short, we could say that it is a theory that puts emphasis on cognition at the service of real interactions with the environment (for a review, see Wilson, 2003). Our ideas and thoughts of the world exist as a result of continuous interactions with our surrounding environment, and our ability to process that information exists for interacting with this environment once again. In conclusion, cognition and perception are not closed and isolated processes: we receive input from the external world while we are interacting with it in a never-ending loop. Beyond physical attributes and genetic factors, different groups of people who have been exposed to different environments might have different cognitive strategies to make sense of the world. This is the reason why **individual differences** between groups of users based on age, gender or cultural background must be taken into consideration when studying cognition for designing purposes. In the specific case of ageing, we also know that some aspects of cognition decline naturally, so identifying them in order to learn how can they affect interaction with a specific system is a good way of starting to take older adults into account.

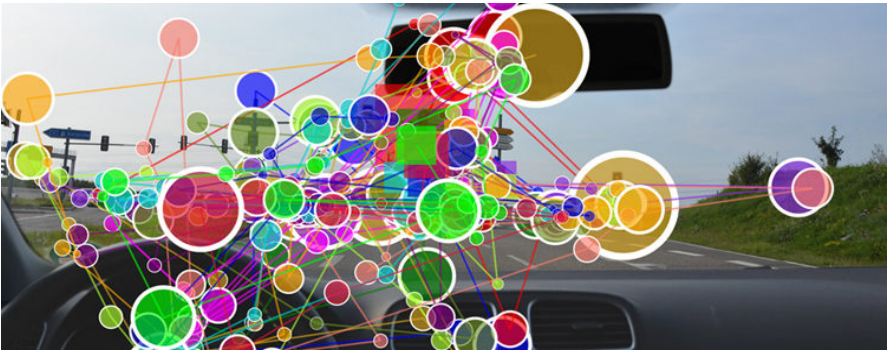

*Figure 2. An example of eye-tracking technique applied to driving, an everyday task. Each circle represents a different fixation (eyes in stationary position), the size of each circle represents the time of each fixation, and the straight lines between fixations represent each saccade (eye movements between fixations). Source: Raschke (2016).*

**Applying cognition to design**

I will use autonomous driving one last time as an example of a CIS older adults can benefit from. Despite autonomous driving being a reality, these systems are not fully ready for total automation and there is a human component that is still very important. This is because the autonomous driving function is not capable of solving every situation and manual control can be required at any point. Some questions that follow could be: What is the best way to indicate that the driver has to regain control? How can we make it sure that the driver does not trust blindly in the system so they do not disengage completely from the supervision task? These questions are not new, but they are investigated with a general population in mind. If we think specifically about older adults, the main logic would be the following:

> Autonomous driving has potential benefits for older adults, for example, enabling them to regain some independence by leaving home more often if they so desire.
>
> Autonomous driving could be particularly dangerous for older adults if their needs are not taken into account, which would also make them unwilling to use these types of systems. This turns a potential chance for independence into something to avoid.
>
> We could investigate what older adults need in this particular context and how to maximise benefits and reduce disadvantages that using these systems may have for them.

In a driving simulator study, Ramkhalawansingh , Keshavarz, Haycock, Shahab and Campos (2016) showed that there are cognitive differences between young drivers and old drivers in the way that they integrate visual and auditive information. Participants of the study were told to drive at 80km/h, but after some time, the speedometer disappeared and they had to maintain that speed. Results showed that older adults were able to maintain their speed more easily than younger drivers when sounds of wind and engine were congruent with the driving image on the screen. Considering these results, one possible next step and question of research applied to autonomous driving could be: Is it possible that older adults would react faster to a disconnection that alerts the driver with a sound and image that are somehow congruent to each other than they would to another type of alert? That is a possibility that might be worth exploring, but this is just one of the infinite examples of how useful it could be to apply cognitive psychology to design, taking into account individual differences.

**Conclusion**

The age increase of the world population is occurring in parallel to the fast-paced development of new interactive technologies. Instead of turning these new systems into a source of frustration and potential danger for older adults, now we have the chance to study these adults' cognitive idiosyncrasies so they can make the best of the new technologies in terms of security and acceptability. Beyond opinion, we need to study those objective factors that humans are not necessarily aware of from the perspective of cognitive psychology. To do this, we must ensure that we experiment with participants under naturalistic and multimodal conditions that are easily generalisable to the type of potential interaction with the system that will be de-signed.

**About the author:**
Lucas Morillo-Mendez is a PhD student at the Center for Applied Autonomous Sensor Systems (AASS) at Örebro University. He obtained his BSc in psychology from Complutense University of Madrid in 2013 and he completed his MSc in Cognitive and Clinical Neuroscience from Goldsmiths College (University of London) in 2014. Lucas's has also worked as a clinical neuropsychologist and in the field of Human Factor research for the automotive industry (Galician Automotive Technology Centre). As a natural consequence of having been surrounded by engineers and stunning new technologies for so long, his academic interests are currently orbiting the interdisciplinary field of Human-Machine Interaction. His current research focuses on studying the impact that certain aspects of visuo-auditory perception have in the design of cognitive interaction systems for different user profiles, especially older adults. Lucas is employed at Örebro University since 2018 as part of the Newbreed-Successful ageing interdisciplinary doctoral program, co-funded by the European Commission through the Marie Skłodowska-Curie Actions (MSCA COFUND).